\documentclass[12pt,a4paper]{article}

\usepackage{comment}
\usepackage{ulem}
\usepackage{tikz}

\usepackage{mathtools}

\usepackage[mathscr]{euscript}

\usepackage{slashed}
\newcommand{\sh}{\slashed}

\usepackage{amsmath}

\date{}

\title{\textbf{Pauli-Villars Regularization Elucidated in
Bopp-Podolsky's Generalized Electrodynamics
}}
\author{ \textbf{
Chueng-Ryong Ji$^{a}$,
Alfredo T. Suzuki$^{b}$,
Jorge H. O. Sales$^{c}$,}
\\\textbf{Ronaldo Thibes$^{d}$}
\\\\
\textit{$^{a}$\small{Department of Physics, North Carolina State University, Raleigh, North Carolina 27695-8202}},\\
\textit{$^{b}$\small{Physics Department, La Sierra University, Riverside, CA 92505}},\\
\textit{$^{c}$\small{Universidade Estadual de Santa Cruz -- DCET, 45662-900 Ilheus, BA -- Brazil}},\\
\textit{$^{d}$\small{Universidade Estadual do Sudoeste da Bahia -- DCEN, Itapetinga, BA -- Brazil}}\\
 }

\usepackage{graphicx}
\usepackage{amssymb,amsfonts,amssymb,amsthm}
\usepackage{amsmath}

\begin{document}

\maketitle

\abstract{
We discuss an inherent Pauli-Villars regularization in Bopp-Podolsky's generalized electrodynamics.
Introducing gauge-fixing terms for Bopp-Podolsky's generalized electrodynamic action,
we realize a unique feature for the corresponding photon propagator with a built-in Pauli-Villars regularization
independent of the gauge choice made in Maxwell's usual electromagnetism.
According to our realization, the length dimensional parameter $a$ associated with Bopp-Podolsky's higher order derivatives corresponds to the inverse of the Pauli-Villars regularization mass scale $\Lambda$, i.e. $a = 1/\Lambda$. Solving explicitly the classical static Bopp-Podolsky's equations of motion for a specific charge distribution, we explore the physical meaning of the parameter $a$ in terms of the size of the charge distribution. As an offspring of the generalized photon propagator analysis, we also discuss our findings regarding on the issue of the two-term vs. three-term photon propagator in light-front dynamics.
}

\section{Introduction}\label{int}
Quantum electrodynamics (QED) may be regarded as a prototype of quantum field theories with well-established renormalization program which effectively regulates the infinities present in the local gauge field theory. Due to the infinities that cannot be gotten around, e.g. radiative corrections in QED, one needs to treat and tame such infinities taking a certain regularization procedure with the renormalization condition for physical amplitudes. The very impressive agreement between high precision measurements in accelerators and the predictions of quantum field theory in the presence of radiative corrections is the key for the indication of successful renormalization program. Phenomenological success of atomic model appears ultimately backed up by the successful QED renormalization program. 


Historically, the problem of infinities first arose in the classical electrodynamics of point particles in the 19th and early 20th century. The well-known example is the mass of electron including the electromagnetic mass $m_{\rm em}$ due to its own electrostatic field given by $m_{em}=\frac{e^2}{8\pi r_e}$ with the charge $e$ and the radius $r_e$ of the electron, which becomes an infinity as $r_e \to 0$. It may not be an overstatement that the early work of Lorentz and Abraham~\cite{AL}
including the bare mass of the spherical shell as well as $m_{\rm em}$ to take a consistent point limit provided the inspiration for later development of the renormalization program in QED and other local field theories. Modifying the concept of point charge to an extended charge distribution lends
also the physical meaning of charge renormalization as the charge screening 
due to the Dirac vacuum in QED. 

In the same vein, Bopp \cite{Bopp} and Podolsky \cite{Podolsky:1942zz} attempted to remove infinities inherent in the usual treatment of point charges introducing higher order derivatives in the Lagrangian of electrodynamics while maintaining the equations of motion still linear in the fields and preserving gauge invariance. In particular, Podolsky discussed the classical aspects of his model presenting the equations of motion, energy-momentum tensor and plane wave field solutions \cite{Podolsky:1942zz}.
Traditionally, however, it has become the case to view the model due to Bopp and Podolsky (``BP model") as a mechanism to describe massive photons without breaking gauge invariance as the propagating modes of the model comprise both massless photons as well as massive ones. In this work, we demonstrate that the BP model solution for a point charge in electrodynamics corresponds to an ordinary electrodynamic solution for a specific charge distribution.  
Motivated by this possible reinterpretation of BP model solution in electrodynamics, we further elucidate the BP model 
as a natural way of providing Pauli-Villars regularization in ordinary QED.  

To discuss our work, it may be worthwhile to make a brief historical remark on previous works on the BP model.
A few years later after the introduction of the BP Lagrangian, Podolsky and Kikuchi \cite{Podolsky:1944zz} 
went through the actual quantization of the model.  They
claimed that the usual quantization methods of the time could not be directly applied
to the BP model of generalized electrodynamics due to the presence of higher order derivatives in the Lagrangian
and therefore they needed to introduce extra auxiliary fields.
The quantization was performed in an extended phase space and, to take gauge symmetry issues into account, a generalization of the
Stueckelberg formalism was used.
A review of BP model's original results by Podolsky and Schwed can be found in Ref. \cite{Podolsky:1948}.
Some forty years later, the BP model was revisited when 
Galv\~ao and Pimentel \cite{Galvao:1986yq} first carefully
analyzed its structure of constraints performing
the instant-form canonical quantization with Dirac Brackets.
In terms of the canonical Dirac-Bergmann formalism \cite{Dirac,Bergmann}, BP model has three first-class constraints generating
gauge symmetries \cite{Galvao:1986yq}.
The canonical quantization was performed after gauge fixing and promotion of Dirac Brackets into operator commutators.
It is worth noting though that in Ref. \cite{Galvao:1986yq} the term of BP model's with higher order derivatives was considered with the opposite sign.
In fact, the electrodynamics of BP model with the opposite sign was then further investigated in a series
of papers \cite{Galvao:1986yq, BarcelosNeto:1991dp, Belvedere:1991xs}, leading to the discussion of tachionic propagating modes for the photon. With the advent of modern and more powerful quantization methods, Barcelos-Neto, Galv\~ao and Natividade \cite{BarcelosNeto:1991dp} performed the
Batalin-Fradkin-Vilkovisky (BFV) quantization using two slightly different gauges, namely the usual Lorenz gauge
and the other called {\it generalized Lorenz gauge}. More recently, Bufalo and Pimentel \cite{Bufalo:2013joa} extended the BFV analysis including matter fields. From the symplectic quantization point of view, it is also worth mentioning that an interesting duality connection between the BP model and the massive Proca model has been investigated in Ref.
\cite{Abreu:2010zza}. More recent further discussions on the BP model can be found in Refs. \cite{Bonin:2009je, Bertin:2009gs, Cuzinatto:2011zz, Zayats:2013ioa, Bufalo:2014jra, Gratus:2015bea, Barone:2016pyy, Thibes:2016ivt}.

In the case of the Lorenz gauge, a natural gauge-fixing term originally introduced by Podolsky and Kikuchi~ \cite{Podolsky:1944zz} which considerably
simplifies the calculation in the quantization process has long been passed without notice in the literature and has been only
recently rescued by Bufalo, Pimentel and Soto \cite{Bufalo:2014jra}.  The role of this term in obtaining a simple generalized photon propagator in a straightforward manner cannot be overemphasized. In particular, we show that this term permits a nice factorization of the generalized photon propagator in all gauges analyzed in this work.
Thus, the BP model parameter dependent part of the
propagator appears only as a global multiplicative factor turning its mathematical structure easier 
to analyze and interpret as a way of introducing Pauli-Villars regularization. 
We discuss that it is possible to split the propagator as a sum of two parts consisting 
of massive and massless modes which we interpret as the Pauli-Villars regularization.

Furthermore, to our knowledge neither the axial nor the light-front
gauges have been discussed so far by the functional path-integral quantization point of view in the context of BP model in the literature.
The canonical structure of BP model's generalized electrodynamics on the null-plane has been recently analyzed by Bertin, Pimentel and Zambrano \cite{Bertin:2009gs}
where the light-front Hamiltonian form evolution is considered.  In Ref. \cite{Bertin:2009gs}, after unraveling the constraint structure in phase space,
the generalized radiation gauge on the null-plane is adopted.
In the present work, we follow a different approach considering the theory defined by the Lagrangian in the configuration space and introducing the gauge-fixing conditions in the
integration measure of the generating functional via the Faddeev-Popov procedure generalized to the BP model case.  The quantization is then performed in a covariant
way and for instance the generalized Lorenz gauge can be achieved.  The axial-gauges are obtained along the same lines, the breaking of relativistic covariance being
only perceived by a particular choice of the axial direction vector $n_\mu$.
In addition, we have the opportunity to extend the ideas introduced in  Refs. \cite{Srivastava:2000cf, Suzuki:2003jz}  concerning the adoption of two simultaneous gauge-fixing conditions leading to a so-called doubly transverse
photon propagator in the light-front gauge. We show in this work explicitly how to handle the corresponding calculations in the BP model case.
   
Our work is organized as follows.
In Section {\bf \ref{pge}}, we define our notation and conventions,
review some basic facts about the BP model as a gauge field theory
and physically interpret some of its classical properties.
In particular, we demonstrate that the BP model solution for a point charge in electrodynamics corresponds to an ordinary electrodynamic solution for a specific charge distribution.  This motivates us in Section {\bf \ref{GF}} to elucidate the BP model as a natural way of providing Pauli-Villars regularization in ordinary QED. 
We discuss in Section {\bf \ref{GF}} the covariant Lorenz gauge fixing 
obtaining the corresponding photon propagator and point out the necessity of the natural gauge-fixing term in the gauge fixing action. This term permits a natural factorization of the generalized photon propagator in all gauges analyzed in Section {\bf \ref{GF}}. The BP model parameter dependent part of the
propagator appears only as a global multiplicative factor turning its mathematical structure easier 
to analyze and interpret as a way of introducing Pauli-Villars regularization. 
In Section {\bf \ref{ese}}, we provide an example of the BP model application discussing
the second-order correction to the electron self-energy and show explicitly the consistency
with the Pauli-Villars regularized result. 
We close in Section {\bf \ref{con}} with some final comments and concluding remarks.
In Appendix A, we summarize the derivation of Eq.(\ref{barpsiabr2}) used in Section {\bf \ref{pge}}. 

\section{Bopp-Podolsky's Generalized Electrostatics}
\label{pge}
The starting classical action in Bopp-Podolsky (BP)'s generalized electrodynamics containing second order space-time derivatives
of the gauge field $A_\mu(x)$ in Minkowski space is given by
\begin{equation}
\label{acao}
 S_0[A_\mu] = \int \,d^4x\,\left\{ 
 -\frac{1}{4}F_{\mu\nu}F^{\mu\nu}
 +\frac{
 a^2
 }{2}\partial_\nu F^{\mu\nu}\partial^\rho F_{\mu\rho}
 \right\}
\end{equation}
where $a$ is a real number with physical dimension of length or inverse mass, known as {Bopp-Podolsky's parameter \cite{Bopp,Podolsky:1942zz}}.
We use Minkowski's coordinates with metric signature diag($\eta^{\mu\nu}$) $=(+1,-1,-1,-1)$ and the integration measure $d^4x$ in Eq.(\ref{acao})
runs throughout all space-time coordinates $x^\mu$.
It is clear that BP's classical action $S_0$ is a natural higher derivatives
Lorentz covariant
generalization of ordinary Maxwell's electromagnetism -- the latter being
recovered for $a=0$.
Although with a different notation, we adopt the same original {Bopp-Podolsky's \cite{Bopp, Podolsky:1942zz, Podolsky:1944zz, Podolsky:1948} choice} for
the second term in Eq.(\ref{acao}).
This choice is important if one wishes to interpret the extra degrees of freedom of BP model as physical massive excitations.
Although the case of negative sign for the second term in $S_0$ above was considered in \cite{Galvao:1986yq, BarcelosNeto:1991dp, Belvedere:1991xs} describing tachionic mass excitations for the gauge field,
we are not going to discuss
it here but rather maintain the implementation consistent with the usual causality.

Note that in Eq.(\ref{acao}), the short-hand $F_{\mu\nu}\equiv\partial_\mu A_\nu - \partial_\nu A_\mu$ stands for the ordinary electromagnetic field strength tensor which is naturally invariant
under
the gauge group $U(1)$.  That means the BP extra $a$-dependent term does not spoil the original gauge invariance of the action $S_0$
and, particularly, the propagator of the gauge field is not well defined before gauge fixing.
We shall address the gauge fixing issue in Section {\bf \ref{GF}} where
the Lorenz and axial type gauge fixings will be discussed.
In the following we briefly review a few immediate properties and consequences of action given by Eq.(\ref{acao}) and 
consider the static case obtaining the BP version of Poisson's equation as well as its general solution.
For a point charge delta distribution, the BP model leads to a everywhere finite potential -- we shall show
that it is possible to generate this very same potential within the scope of
ordinary electrodynamics using a suitable charge distribution.

\subsection{Field Equations of Motion and General Static Solution}
In order to understand the physical content encoded in Eq.(\ref{acao})
we couple the gauge field to an external source $j^\mu(x)$ through
\begin{equation}
S_{ext}=-\int d^4x\, j^\mu A_\mu
\end{equation}
and demand stationarity of the total action under functional
variations of the gauge field $A_\mu$.
As a result, the corresponding Euler-Lagrange equations of motion
obtained from the minimal action principle read
\begin{equation}\label{EL}
(1+a^2\square)\partial_\mu F^{\mu\nu}=j^\nu\,,
\end{equation}
and
can be understood in the present context as Maxwell's generalized equations.
As the indices $\mu, \nu$ run from $0$ to $3$, Eq.(\ref{EL})
represents a set of four fourth-order partial differential equations on the gauge field $A_\mu$.
Conservation of the external current $j^\mu$ comes straightforwardly from the antisymmetry of the field strength
\begin{equation}
\partial_\nu j^\nu = (1+a^2\square)\partial_\nu \partial_\mu F^{\mu\nu} \equiv 0\,.
\end{equation}
In terms of the measurable usual electric and magnetic fields respectively
given by
\begin{equation}\label{Ei}
E^i=-F^{0i}\,,
\end{equation}
and
\begin{equation}\label{Bi}
B^i=-\frac{1}{2}\epsilon^{ijk}F^{jk}\,,
\end{equation}
the action given by Eq.(\ref{acao}) can also be expressed as
\begin{equation}
S_0=\frac{1}{2}\int d^4x\,
\left[
{\bf E}^2-{\bf B}^2 +a^2({\bf\nabla} \cdot {\bf E})^2 - a^2({\dot {\bf E}}-{\bf\nabla}\times{\bf B})^2
\right]
\,,
\end{equation}
while Eq.(\ref{EL})
gives rise to the four non homogeneous Maxwell equations, namely, one for the temporal component
\begin{equation}\label{tcomp}
(1+a^2\square) {\bf \nabla}\cdot{\bf E} = j^0
\,,
\end{equation}
and three for the space components
\begin{equation}\label{scomp}
(1+a^2\square)({\bf \nabla}\times{\bf B}-\frac{\partial{\bf E}}{\partial t})={\bf j}
\,. 
\end{equation}
The homogeneous Maxwell equations, on the other hand, remain the same as they amount to the very identities which
permit to describe the physical $\bf E$ and $\bf B$ fields, through Eqs. (\ref{Ei}) and (\ref{Bi}) above, in terms
of the gauge potential field $A_\mu$. In particular the absence of magnetic
monopoles still holds in BP's generalization as the magnetic field $\bf B$ remains
divergenceless.

Let's now focus on the static case where it is possible to obtain the general solution and discuss the physical meaning of the BP model. 
For a time-independent electromagnetic field, Eqs. (\ref{tcomp}) and (\ref{scomp}) reduce respectively to
\begin{equation}\label{ggl}
(1-a^2{\bf \nabla}^2) {\bf \nabla}\cdot{\bf E} = j^0
\end{equation}
and
\begin{equation}
(1-a^2{\bf \nabla}^2) {\bf \nabla}\times{\bf B} ={\bf j}
\end{equation}
containing now only space derivatives,
while the homogeneous ones simply state that $\bf B$ is divergenceless and $\bf E$ irrotational.  In this case,
inserting (\ref{Ei}) disregarding time derivatives into Eq.(\ref{ggl}) leads to a
generalized Poisson equation
\begin{equation}\label{gPe}
(1-a^2{\bf \nabla}^2){\bf \nabla}^2\phi=-4\pi\rho \, ,
\end{equation}
where $\phi\equiv A_0$ represents the electrostatic potential
and we have written $j^0=4\pi\rho$ for the electric charge density.
By defining the $a$-dependent fourth-order differential operator ${\mathbb P}_a$ as
\begin{equation}
 {\mathbb P}_a \equiv (1-a^2{\bf \nabla}^2){\bf \nabla}^2
\end{equation}
we may rewrite Eq.(\ref{gPe}) more compactly as
\begin{equation}\label{PP}
 {\mathbb P}_a \phi=-4\pi\rho
\,.
\end{equation}
Henceforth, we shall refer Eq.(\ref{PP}) as the Poisson-Bopp-Podoslky (PBP) equation.

To solve the PBP equation given by Eq.(\ref{PP}), equivalently Eq.(\ref{gPe}), for an arbitrary given charge distribution $\rho$, let us first consider the case of an elementary distribution resulting from a fixed unity point
charge localized at $\mathbf{r}_0$
\begin{equation}\label{rhoC}
 \rho_{\mathbf{r}_0}(\mathbf{r}) = \delta^{(3)}(\mathbf{r}-\mathbf{r}_0)
\,.
\end{equation}
For this case the solution $\phi_{P,a}(r)$ can be written as the difference between the usual Coulomb potential
\begin{equation}\label{Coulomb}
 \phi_C(r) \equiv \frac{1}{r}
\end{equation}
and the Yukawa potential
\begin{equation}\label{phi_Y}
 \phi_{Y,a}(r) \equiv \frac{e^{-r/a}}{r}
\end{equation}
evaluated at $r = |\mathbf{r}-\mathbf{r}_0|$.
In fact, the BP potential centered at $\mathbf{r}_0$, defined as
\begin{equation}\label{phi_P}
 \phi_{P,a}(|\mathbf{r}-\mathbf{r}_0|) \equiv \phi_C(|\mathbf{r}-\mathbf{r}_0|) - \phi_{Y,a}(|\mathbf{r}-\mathbf{r}_0|)
 = \frac{ 1 - e^{-|\mathbf{r}-\mathbf{r}_0|/a} }{|\mathbf{r}-\mathbf{r}_0|}
\,,
\end{equation}
satisfies Eq.(\ref{PP}) for the elementary Dirac delta charge distribution given by Eq.(\ref{rhoC}).  This can be directly seen by applying the
${\mathbb P}_a$ operator to each potential leading to
\begin{equation}\label{19}
 {\mathbb P}_a \phi_C (\mathbf{r}) =
 -4\pi \left[
 \delta^{(3)}(\mathbf{r})-a^2{\bf \nabla}^2\delta^{(3)}(\mathbf{r})
 \right]
\end{equation}
and
\begin{equation}\label{20}
 {\mathbb P}_a \phi_{Y,a} (\mathbf{r}) = 4\pi a^2{\bf \nabla}^2\delta^{(3)}(\mathbf{r})
\,,
\end{equation}
and then subtracting Eq.(\ref{20}) from Eq.(\ref{19}).

Applying the Green's function method, we then find that the general solution $\phi(\mathbf{r})$ of the PBP equation given by Eq.(\ref{PP}) subject to the boundary condition of vanishing potential at infinity can be written as a superposition
of kernel elementary contributions (\ref{phi_P}) weighed by the given charge density $\rho(\mathbf{r})$, that is
\begin{equation}\label{gensol}
 \phi({\mathbf r}) = \int \rho({\mathbf r}')\frac{ \left(1 - e^{-|\mathbf{r}-\mathbf{r}'|/a} \right) }{|\mathbf{r}-\mathbf{r}'|} d\tau'
\, ,
\end{equation}
where $d\tau'$ denotes the integration volume element with respect to the dummy integration variable $\mathbf{r}'$.
The ordinary electrostatic solution is recovered in the limit that the BP model parameter $a$ goes to zero, i.e. $a\rightarrow 0$, as expected. 

\begin{figure}
  \centering
  \includegraphics[width=8cm]{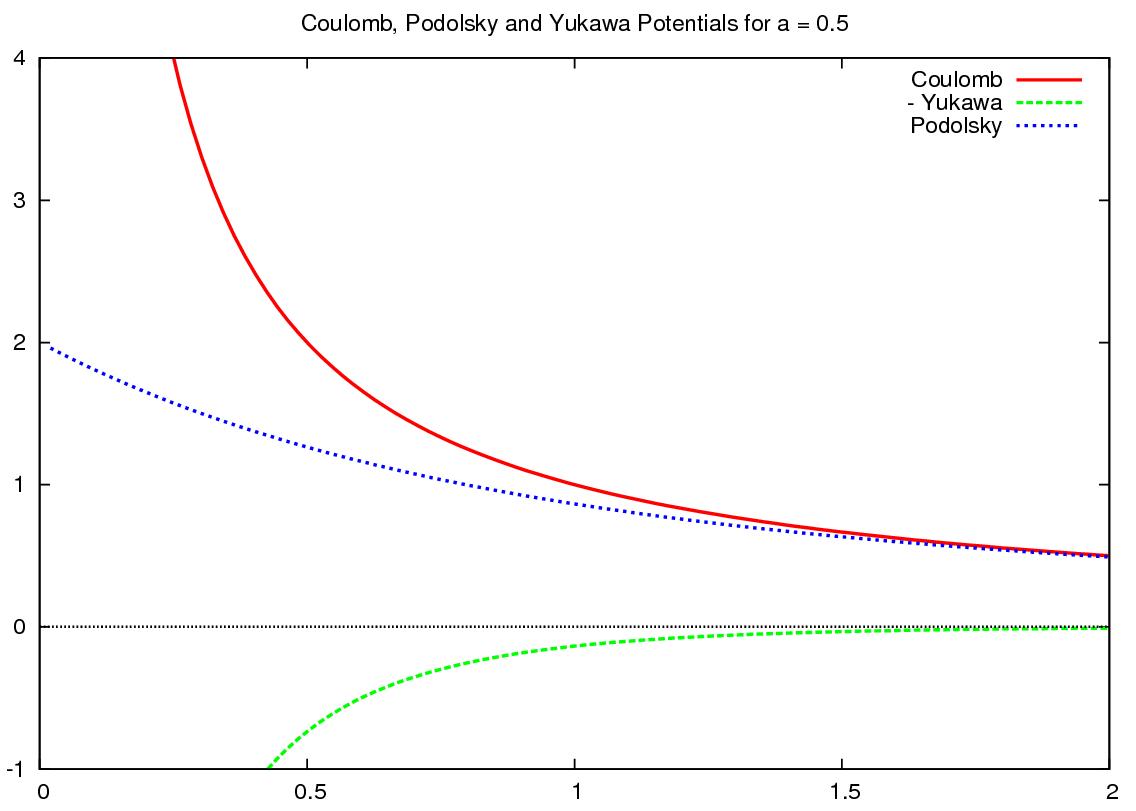}
  \caption{Podolsky potential for $a=0.5$ as a function of the distance.  Although both Coulomb's and Yukawa's potentials blow up at the origin,
their difference remains finite. In the limit $a\rightarrow 0$ Podolsky's potential bends continually towards Coulomb's potential.}
  \label{fig1}
\end{figure}

For the point charge density given by Eq.(\ref{rhoC}), the ordinary electrostatic solution given by the Coulomb electrostatic potential $\phi_C (r) = \frac{1}{r}$ in Eq.(\ref{Coulomb}) diverges at the local point  $\mathbf{r} = \mathbf{r}_0$, i.e. $r = |\mathbf{r} - \mathbf{r}_0| = 0$, imposing the problem of infinities discussed in our Introduction, Sec.\ref{int}, for the classical electrodynamics of point particles. On the other hand, for non-null $a$, the BP potential $\phi_{P,a}(r)$ 
in Eq.(\ref{phi_P}) remains finite in the limit $r\rightarrow 0$ approaching to the finite value
\begin{equation}
 \phi_{P,a}(0) = \frac{1}{a}
 \,
\end{equation}
and reproduces back the Coulomb's characteristic $1/r$ behavior for large values of $r$ compared to $a$.
In Fig.\ref{fig1},
we plot BP's potential 
as a function of $r$
as well as its two constituent parts Coulomb's and minus Yukawa's for the numerical value $a=0.5$
in the same unity as the measured distance variable $r$. Here, it may be worth noticing
the important fact that Coulomb's potential doesn't have any length scale parameter while the BP's potential has a natural length scale provided by the real parameter $a$. What follows in the next subsection is that this parameter $a$ 
introduced in the BP model action given by Eq.(\ref{acao}) can be equivalently reinterpreted as the length scale of a specific charge distribution removing the fiasco of divergence for a point charge particle in ordinary local 
gauge electrodynamics. In particular, we demonstrate that the BP model solution for a point charge corresponds to an ordinary electrodynamic solution for a specific charge distribution. This motivates us in the subsequent section, Sec. {\bf \ref{GF}}, to elucidate the BP model as a natural way of providing Pauli-Villars regularization in ordinary QED. 

\subsection{The BP Potential from Ordinary Electrostatics}
As discussed in our Introduction, Sec.\ref{int}, a non-null BP $a$-parameter improves the model convergence properties by means of the higher-order derivatives term.
As a matter of fact we have just seen Coulomb's potential given by Eq.(\ref{Coulomb})
gets smoothed out becoming finite at the critical point $\mathbf{r}=\mathbf{r}_0$
when we generalize Poisson's equation including the convergence parameter
$a$ into the PBP equation given by Eq.(\ref{gPe}) or Eq.(\ref{PP}).
In other words, in the BP model, Eq.(\ref{Coulomb}) generalizes to Eq.(\ref{phi_P}).
However, we show below that it is also possible to obtain the same effect within the realm
of ordinary electrostatics by using a specific charge distribution.
Indeed, let's take a small positive length dimension parameter $b$ and
consider the normalized charge distribution
\begin{equation}\label{rho2}
\rho(b,r)=\frac{e^{-r/b}}{4\pi b^2r}
\,.
\end{equation}
Then, the general solution to Poisson's equation gives the corresponding potential
\begin{equation}\label{V}
V(b,r) = \frac{1}{4\pi b^2} \int\,d\tau'\, \frac{e^{-r'/b}}{r'|\mathbf{r}-\mathbf{r}'|}
\,.
\end{equation}
Using the spherical coordinates and writing
\begin{equation}
d\tau'= r'^2\sin\theta \,d\phi \,d\theta \,dr' ,
\end{equation}
one can see straightforwardly that the azimuthal part in $\phi$ brings a $2\pi$ contribution and the integral in Eq.(\ref{V})
is directly worked out
as
\begin{eqnarray}
V(b,r) &=& \frac{1}{2b^2}\int_0^\infty
dr'\,r'e^{-r'/b}\int_0^\pi \frac{d\theta\sin\theta}{\sqrt{r^2+{r'}^2-2rr'\cos \theta'}}
\,,
\nonumber\\
&=&\frac{1}{b^2}\int_0^\infty \frac{dr'\,e^{-r'/b}}{r}\displaystyle\left\{ |r+r'|-|r-r'| \right\}
\,,
\nonumber\\
&=&\frac{1}{b^2}
\int_0^r dr'\, \frac{r' e^{-r'/b}}{r} + \frac{1}{b^2}\int_r^\infty dr'\,e^{-r'/b} 
\,.
\end{eqnarray}
Performing then the last two $r'$ integrations, we obtain finally
\begin{equation}\label{BPb-potential}
V(b,r)=\frac{1-e^{-r/b}}{r}\equiv\phi_{P,b}(r)
\,.
\end{equation}
Note the key fact that we dispensed completely
the BP higher-derivative $a$-dependent term in obtaining Eq.(\ref{BPb-potential}) 
as we used only the standard ordinary electrodynamics. Nevertheless, it represents the very BP potential with
$b$ playing the role of the previous BP $a$-parameter. 

To discuss this key point further, let's now address the following question: i.e.,\\
Which potential would be produced by the charge distribution given by Eq.(\ref{rho2}) in
the BP model context using the general solution given by Eq.(\ref{gensol})?  
The answer to this question turns out to be an $(a,b)$-dependent potential which we denote here by $\psi(a,b,r)$.
This sort of double BP potential can be explicitly calculated using the general solution given by (\ref{gensol}) as
\begin{equation}\label{psiabr}
 \phi({\mathbf r}) = \frac{1}{4\pi b^2}\int \frac{ e^{-r'/b}\left(1 - e^{-|\mathbf{r}-\mathbf{r}'|/a} \right) }{r'\,|\mathbf{r}-\mathbf{r}'|} d\tau' .
\end{equation}
The first part of Eq.(\ref{psiabr}) has been previously calculated and shown
to represent $\phi_{P,b}(r)$.  Therefore, we write
\begin{equation}\label{psiabr2}
\psi(a,b,r) = \phi_{P,b}(r) - \bar\psi(a,b,r)
\end{equation}
with
\begin{equation}\label{barpsiabr}
\bar\psi(a,b,r)\equiv	
\frac{1}{4\pi b^2}\int \,d\tau'\,\,
\frac{ \exp\left[-\frac{r'}{b}-\frac{|\mathbf{r}-\mathbf{r}'|}{a}\right]}{r'|\mathbf{r}-\mathbf{r}'|}\,\,.
\end{equation}
The integration of Eq.(\ref{barpsiabr}) is explicitly given in appendix \ref{A}, which shows
that it leads to
\begin{equation}\label{barpsiabr2}
\bar\psi(a,b,r)=
-\frac{e^{-{r\over b}} - e^{-\frac{r}{a}}}{r[1-(b/a)^2]}\,.
\end{equation}
By plugging this result back into Eq.(\ref{psiabr2}), we obtain the final result
for the double BP potential as
\begin{equation}\label{abPotential}
\psi(a,b,r)={1\over r} +
\frac{a^2e^{-r/a}-b^2e^{-r/b}}{r(b^2-a^2)}
\,.
\end{equation}
Specifically for $r$ close to zero, we get a finite result without any divergence:
\begin{equation}
\lim_{r\rightarrow0}\psi(a,b,r) = \frac{1}{a+b}
\,.
\end{equation}
We note here that in the $a\rightarrow0$ limit, Eq.(\ref{abPotential})
reproduces Eq.(\ref{BPb-potential}), or vice-versa, i.e. in the $b\rightarrow0$ limit, Eq. (\ref{abPotential})
reproduces Eq.(\ref{phi_P}). 
The double BP potential given by Eq.(\ref{abPotential}) is completely symmetric with respect to $a$ and $b$ and 
reveals the equivalence of the result under the exchange of the role between 
the two parameters $a$ and $b$ which have been introduced originally with seemingly different physical 
motivation or physical meaning. The result of the ordinary electrostatics, i.e. $a=0$, for a specific charge charge distribution given by Eq.(\ref{rho2}) with $b \neq 0$ is completely equivalent to the result of the BP electrostatics
with a length scale parameter $a \neq 0$ for a local point charge, i.e. $b=0$. Thus, it allows the exchange of the role
between the length scale parameter $a$ introduced in the BP model for a point charge and the length scale parameter
$b$ for a specific charge distribution given by Eq.(\ref{rho2}) in the ordinary electrostatics.  
To the extent that modifying the concept of point charge to an extended charge distribution provides
the physical meaning of charge renormalization as the charge screening due to the Dirac vacuum in QED,
our finding here motivates us to reinterprete the BP's generalized electrodynamic action
given by Eq.(\ref{acao}) as a natural way of providing Pauli-Villars regularization in ordinary QED.
In the next section, we address this possible reinterpretation by looking into the details of the gauge fixing 
in the BP model with its functional quantization.

\section{Gauge Fixing and Functional Quantization}\label{GF} 
As already stated in the last paragraph of our Introduction, Sec.1, due to gauge invariance, a direct propagator for the
gauge field in BP model is ill-defined. That happens because we are working with a constrained system
and to preserve explicit covariance we use more field variables than degrees of freedom \cite{Thibes:2016ivt}.  In order to proceed with
the quantization of the model, similarly to ordinary electrodynamics,
we must choose a specific gauge suitable for perturbative calculations.
In the following subsections, we show how to achieve the generalized Lorenz and axial gauges performing the functional quantization of BP model.
In both cases, we shall obtain the Green functions generating functional
by means of a suitable generalization of the Faddeev-Popov method.
\subsection{Generalized Lorenz Gauge}\label{lgf}
Concerning the covariant Lorenz gauge we generalize the standard gauge-fixing term in the Maxwell Lagrangian to
\begin{equation}\label{LGF}
 {\cal L}_{LGF}=-\frac{1}{2\xi}
 (\partial_\mu A^\mu)^2
 +\frac{a^2}{2\xi}(\partial_\lambda\partial_\mu A^\mu)(\partial^\lambda\partial_\nu A^\nu)
\end{equation}
which is then added to the original Lagrangian in the integrand of action given by Eq.(\ref{acao}).
Besides BP's $a$-parameter, the gauge-fixing given by Eq.(\ref{LGF}) is allowed to depend on the additional free real gauge parameter $\xi$.
For the particular case $\xi=1$ the necessity of this natural term can already be seen in the original papers 
of Podolsky, Kikuchi and Schwed \cite{Podolsky:1944zz,Podolsky:1948}.  
Although this isolated term cannot be found directly in Ref.\cite{Podolsky:1944zz}, a careful reading shows that it is in fact inserted and summed up in their so called
{\it modified Lagrangian} up
to total divergences.
However, the second part of Eq.(\ref{LGF}) has been tacitly omitted in the more modern
treatments of BP's model
and only recently has it been reintroduced in BP's context by Bufalo, Pimentel and Soto \cite{Bufalo:2014jra}.

With the addition of (\ref{LGF}) the fixed action
$S_{LGF}=S_0+\int d^4x\,{\cal L}_{LGF}$ now reads
\begin{equation}\label{caction}
 S_{LGF} =
\int \,d^4x\,\left\{ 
 -\frac{1}{4}F_{\mu\nu}F^{\mu\nu}
 +\frac{
 a^2
 }{2}\partial_\nu F^{\mu\nu}\partial^\rho F_{\mu\rho}
 -\frac{1}{2\xi}
 (\partial_\mu A^\mu)^2
 +\frac{a^2}{2\xi}(\partial_\lambda\partial_\mu A^\mu)^2
 \right\}\,.
\end{equation}
By demanding stationarity with respect to the gauge field, that is, by enforcing
\begin{equation}
 \frac{\delta S_{LGF}}{\delta A_\mu (x)} = 0 \, ,
 \end{equation}
we obtain the equations of motion
\begin{equation}\label{nem}
 (1+a^2\Box)
 \left(
 \partial_\nu F^{\nu\mu} + \frac{1}{\xi}\partial^\mu\partial_\nu A^\nu
 \right)
 = 0\,,
\end{equation}
which generalize Eq.(\ref{EL}) in the case of null external source.  In terms of the gauge field $A_\mu$ 
the equations of motion given by Eq.(\ref{nem}) 
can also be rewritten as
\begin{equation}\label{LGFEM}
  (1+a^2\Box)
  \left(
  \Box\eta^{\mu\nu}-\partial^\mu\partial^\nu+\frac{1}{\xi}\partial^\mu\partial^\nu
  \right)
  A_\nu=0\, .
\end{equation}
For the Feynman-t'Hooft gauge choice $\xi=1$, it reduce to the simpler result
\begin{equation}\label{WE}
 (1+a^2\Box)\Box A_\mu = 0\,.
\end{equation}

This means that in the Lorenz gauge the free field equations of motion comprise a superposition of
plane waves describing both massive and massless particles. 
As previously mentioned we remark the importance of the sign choice in Eq.(\ref{acao}) for this physical
interpretation.
In fact the general solution to Eq.(\ref{WE}) can be written as
\begin{equation}
A_\mu(x) = \int \frac{d^4 k}{(2\pi)^3} 
\left\{
\left[
\delta(k^2)a_\mu(k)+\delta(k^2-1/a^2)b_\mu(k)
\right]
e^{-ik\cdot x} + h.c. 
\right\}
\,,
\end{equation}
for arbitrary functions $a_\mu(k)$ and $b_\mu(k)$.

The generalized photon propagator of the BP model in the current working Lorenz gauge can be read directly from the inverse of
the differential operator acting on $A_\mu$ in Eq.(\ref{LGFEM}). Or equivalently,
by performing
integrations by parts and discarding boundary terms, the action $S_{LGF}$ in Eq.(\ref{caction}) may be
rewritten as
\begin{equation}\label{14}
 S_{LGF}=\frac{1}{2} \int d^4x A_\mu
(1+a^2\square)
 \left[
  \Box\eta^{\mu\nu}-(1-\frac{1}{\xi})\partial^\mu\partial^\nu
 \right]
 A_\nu
 \,.
\end{equation}
In momentum space, this gives rise to the operator
\begin{equation}
 M^{\mu\nu}(k)=-(1-a^2k^2)
 \left(
 k^2\eta^{\mu\nu}-(1-\frac{1}{\xi})k^\mu k^\nu
 \right)
 \,,
\end{equation}
and permits us to write the photon propagator as
\begin{equation}\label{PPP}
 P_{\mu\nu}(k)=
 \frac{-1}{(1-a^2k^2)k^2}
 \left[
 \eta_{\mu\nu}+(\xi-1)\frac{k_\mu k_\nu}{k^2}
 \right]
 \,.
\end{equation}
One can straightforwardly check that $P_{\mu\lambda}M^{\lambda\nu}=\delta_\mu^\nu$,
confirming that Eq.(\ref{LGF}) leads to a neat simple expression for the photon propagator above with
the additional massive simple pole $1/a^2$. The Landau gauge result is obtained for $\xi = 0$.

In the following, we show that
the gauge-fixing term given by Eq.(\ref{LGF}) can be obtained from the original BP Lagrangian by imposing the condition
\begin{equation}\label{cond}
(1+a^2\square)\partial_\mu A^\mu = 0
\end{equation}
in the Green's function generating functional by means of the well-known Faddeev-Popov procedure \cite{Faddeev:1967fc}.
This can be done introducing Eq.(\ref{cond}) via a Dirac delta functional in the integration measure.
Explicitly we
can write the generating functional as
\begin{equation}
Z[j^\mu] = N \int DA_\mu \Delta_{FP}
\,
\delta \left((1+a^2\square)\partial^\mu A_\mu\right)\,\exp\left\{iS_0 - i\int d^4x\, j^\mu A_\mu  \right\}
\end{equation}
where 
$N$ is a normalization constant and
$\Delta_{FP}$ represents the determinant which arises from
the Jacobian of a gauge transformation in the condition given by Eq.(\ref{cond}), that is,
\begin{equation}
\Delta_{FP}=\det \, (1+a^2\square)\square  \,.
\end{equation}
By means of introducing a pair of anticommuting ghost fields $(C,{\bar C})$ and an auxiliary Nakanishi-Lautrup \cite{Nakanishi:1966zz,Lautrup:1967zz}
field $B$, we can rewrite the generating functional, after a convenient redefinition of the normalization factor, as
\begin{eqnarray}
Z[j^\mu]&=&N\int DA_\mu DC D{\bar C} DB
\exp\left\{
iS_0
\right.\nonumber\\&&\left.
+i\int d^4x 
\left[
{\bar C}(1+a^2\square)\square C
+
B(1+a^2\square)\partial^\mu A_\mu
\right.\right.\nonumber\\&&\left.\left.
-\frac{a^2\xi}{2}\partial_\mu B\partial^\mu B
+\frac{\xi B^2}{2}
-j^\mu A_\mu
\right]
\right\}
\,.
\end{eqnarray}
Note that the action in the exponential argument is invariant under the BRS transformation
\begin{equation}
\delta_B A_\mu = \partial_\mu C\,,\,\,\,\,\,\,\delta_B {\bar C}=-B\,,
\end{equation}
where the BRS operator $\delta_B$ has Grassmann parity one.  

A functional integration over the $B$ field finally leads to
\begin{equation}
Z[j^\mu]=N\int DA_\mu DC D{\bar C}
\exp\left\{
i\left[
S_{LGF}+S_{LGG}+S_{ext}
\right]
\right\}
\end{equation}
with $S_{LGF}$ given by Eq.(\ref{caction}),
\begin{equation}
S_{LGG} = \int d^4x\,
{\bar C}(1+a^2\square)\square C
\end{equation}
and
\begin{equation}
S_{ext}=\int d^4x\,j^\mu A_\mu
\,,
\end{equation}
thus justifying the Lorenz gauge fixing term given by Eq.(\ref{LGF}).

We have explicitly shown how the condition given by Eq.(\ref{cond}) leads through the Faddeev-Popov procedure to the gauge fixing term given by Eq.(\ref{LGF}) and calculated the corresponding propagator for BP's generalized electrodynamics.  In the next subsection, we turn our attention to the axial and light-front gauges.

\subsection{Axial and Light-Front Gauges}\label{alc}
The axial gauge fixing has been 
originally introduced by Kummer \cite{Kummer:1961} and since then has been
studied in the literature for a long time.
It is a noncovariant gauge in the sense that it relies on a choice of an arbitrary fixed direction in space-time $n_\mu$.
It encompasses the light-front gauge as a special case when $n_\mu$ is light-like.
For an interesting and lively review of the axial, light-front as well as other noncovariant gauges in the context of non Abelian theories we cite Ref.\cite{Leibbrandt:1987qv}.
In the present case of BP model, in order to implement the axial gauge fixing we pick up a specific constant non-null four-vector $n^\mu$ in
space-time and write 
\begin{equation}\label{LF}
{\cal L}_{LF}=-\frac{1}{2\alpha}(n_\mu A^\mu)^2
+\frac{a^2}{2\alpha}(n_\mu\partial_\lambda A^\mu)(n_\nu\partial^\lambda A^\nu)
\,,
\end{equation}
where $\alpha$ stands for a free gauge parameter.
In ordinary electrodynamics the axial gauge fixing term contains only the first part of Eq.(\ref{LF}).
The second part constitutes the natural generalization for BP's electrodynamics.
According to the nature of the directional vector $n_\mu$ we have different types of axial gauges.  Namely temporal axial, light-front axial and
space or proper axial for respectively timelike, lightlike or spacelike $n_\mu$.  The differences among them lead to important subtleties and turn out to be a key point
in the canonical quantization when one needs clearly to pick up a time direction for the Hamiltonian evolution.  
In our present discussion however, we limit ourselves to the Lagrangian analysis proposing Eq.(\ref{LF}) for a general axial gauge.
Addition of the space-time integral of ${\cal L}_{LF}$ to BP's action given by Eq.(\ref{acao}) results in the axial gauge fixed action
\begin{equation}
 S_{LF} =
\int \,d^4x\,\left\{ 
 -\frac{1}{4}F_{\mu\nu}F^{\mu\nu}
 +\frac{
 a^2
 }{2}\partial_\nu F^{\mu\nu}\partial^\rho F_{\mu\rho}
 -\frac{1}{2\alpha}(n_\mu A^\mu)^2
 +\frac{a^2}{2\alpha}(n_\mu\partial_\lambda A^\mu)^2
 \right\}\,,
\end{equation}
which, after discarding surface integration terms can be recast into
\begin{equation}
S_{LF} = \frac{1}{2}\int d^4x\,
A_\mu
(1+a^2\square)
\left[
\square\eta^{\mu\nu}-\partial^\mu\partial^\nu-\frac{1}{\alpha}n^\mu n^\nu
\right]
A_\nu
\,.
\end{equation}
By performing a Fourier transformation, this action can be rewritten in momentum space as
\begin{equation}
S_{LF}=
\frac{1}{(2\pi)^4}
\int d^4k
{\tilde A}_\mu(k) M^{\mu\nu}(k) {\tilde A}_\nu(-k)
\end{equation}
with
\begin{equation}\label{MLF}
M^{\mu\nu}(k)\equiv-(1-a^2k^2)(k^2\eta^{\mu\nu}-k^\mu k^\nu + \frac{1}{\alpha}n^\mu n^\nu)
\,.
\end{equation}
The photon propagator of the BP model in the light-front gauge is the inverse of $M^{\mu\nu}$, reading explicitly
\begin{equation}
P_{\mu\nu}(k)=\frac{-1}{k^2(1-a^2k^2)}
\left[
\eta_{\mu\nu}
+\frac{(\alpha k^2 + n^2)}{(n\cdot k)^2} k_\mu k_\nu
-\frac{1}{(n\cdot k)}(k_\mu n_\nu + k_\nu n_\mu)
\right]
\,.
\end{equation}
Consistently, for the case $a=0$ this result reduces to the usual photon propagator in the axial gauge \cite{Leibbrandt:1983pj}.
Note that despite the last term in Eq.(\ref{MLF}), a term proportional to $n_\mu n_\nu$ does not show up in the propagator $P_{\mu\nu}$.
For the light-front gauge, we have $n^2 = 0$ and by choosing the gauge parameter $\alpha=0$ we get the simpler expression~\cite{two-term}
\begin{equation}\label{PPLF}
P_{\mu\nu}=
\frac{-1}{k^2(1-a^2k^2)}
\left[
\eta_{\mu\nu}
-\frac{1}{(n\cdot k)}(k_\mu n_\nu + k_\nu n_\mu)
\right]
\,.
\end{equation}

Similarly to the Lorenz gauge, here we can justify the term given by Eq.(\ref{LF}) by imposing the gauge condition
\begin{equation}\label{cond2}
(1+a^2\square)n_\mu A^\mu = 0
\end{equation}
in the generating functional by means of the Faddeev-Popov procedure.  In fact, we have here the Faddeev-Popov determinant
\begin{equation}
\Delta_{FP} = \det (1+a^2\square)n_\mu \partial^\mu
\end{equation}
which can be exponentiated by means of the introduction of a pair of anticommuting ghost fields $(C,{\bar C})$
leading to the generating functional
\begin{eqnarray}
Z[j^\mu]&=&N\int DA_\mu DC D{\bar C} DB
\exp\left\{
iS_0
\right.\nonumber\\&&\left.
+i\int d^4x 
\left[
{\bar C}(1+a^2\square)n_\mu \partial^\mu C
+
B(1+a^2\square) n^\mu A_\mu
\right.\right.\nonumber\\&&\left.\left.
-\frac{a^2\alpha}{2}\partial_\mu B\partial^\mu B
+\frac{\alpha B^2}{2}
-j^\mu A_\mu
\right]
\right\}
\,,
\end{eqnarray}
where $B$ is the Nakanishi-Lautrup field.
Paralleling the Lorenz gauge case, here we also have the BRS symmetry
\begin{equation}
\delta_B A_\mu = \partial_\mu C\,,\,\,\,\,\,\,\delta_B {\bar C}=-B\,.
\end{equation}

Integration over the Nakanishi-Lautrup field leads to an effective action in the exponential argument as
\begin{equation}
S_{ef}=
S_0+\int d^4x 
\left\{
-\frac{1}{2\alpha}(n_\mu A^\mu)^2
+\frac{a^2}{2\alpha}(n_\mu\partial_\lambda A^\mu)(n_\nu\partial^\lambda A^\nu)
+{\bar C}(1+a^2\square) n_\mu \partial^\mu C
\right\}
\end{equation}
showing clearly the appearance of the proposed term given by Eq.(\ref{LF}).

In the usual Maxwell case,
there is a well-known discussion in the literature regarding the propagator of the gauge field in the light-front.
Recently it has been shown \cite{Suzuki:2003jz} that it is possible to consider a mixing of the Lorenz and light-front gauge fixings
leading to the three-term photon propagator \cite{Srivastava:2000cf}.  In the following we show that there exists a natural
generalization of the ideas discussed in Ref.\cite{Suzuki:2003jz} to the current BP model.
Specifically, in order to obtain a three-term propagator, we define the axial Lorenz 
gauge-fixing Lagrangian density as
\begin{equation}
\mathcal{L}_{AL}=-\frac{1}{\beta }(n\cdot A)(\partial \cdot A)+\frac{a^{2}
}{\beta }(n_{\mu }\partial _{\lambda }A^{\mu })(\partial _{\nu }\partial
^{\lambda }A^{\nu })  \label{LLF}
\,,
\end{equation}%
where now we denote by $\beta$ the gauge free parameter.
Proceeding analogously to the previous sections, we integrate Eq.(\ref{LLF}) in space-time and add
the result to the gauge action given by Eq.(\ref{acao}).  After disregarding frontier surface terms, the total gauge fixing action now reads
\begin{equation}\label{SLLF}
S_{AL}=\frac{1}{2}\int d^{4}x\,A_{\mu }\left[ (1+a^{2}\square )(\square
\eta ^{\mu \nu }-\partial ^{\mu }\partial ^{\nu }-\frac{1}{\beta }(n^{\mu
}\partial ^{\nu }-n^{\nu }\partial ^{\mu })\right] A_{\nu }  
\,.
\end{equation}

By inverting the differential operator defined in Eq.(\ref{SLLF}) in momentum space, similarly to the previous cases, we obtain
the generalized photon propagator of the BP model in the axial Lorenz gauge as
\begin{eqnarray}
P_{\mu \nu }\left( k\right)  &=&\frac{-1}{k^{2}(1-a^{2}k^{2})}\left[ \eta_{\mu
\nu }+\frac{\beta ^{2}k^{2}+n^{2}}{(n\cdot k)^{2}-n^{2}k^{2}}k_{\mu }k_{\nu
}-\frac{n\cdot k+i\beta k^{2}}{(n\cdot k)^{2}-n^{2}k^{2}}k_{\mu }n_{\nu
}\right. +  \nonumber \\
&&-\left. \frac{n\cdot k+i\beta k^{2}}{(n\cdot k)^{2}-n^{2}k^{2}}k_{\nu
}n_{\mu }+\frac{k^{2}}{(n\cdot k)^{2}-n^{2}k^{2}}n_{\mu }n_{\nu }\right] 
\label{PROPLF}
\end{eqnarray}

Going back to the particular light-front gauge case where
$n^2=0$ and choosing the gauge parameter $\beta=0$ we get
\begin{equation}\label{3tp}
P_{\mu \nu }\left( k\right) =\frac{-1}{k^{2}(1-a^{2}k^{2})}\left[ \eta_{\mu \nu
}-\frac{1}{(n\cdot k)}(k_{\mu }n_{\nu }+k_{\nu }n_{\mu })+\frac{k^{2}}{%
(n\cdot k)^{2}}n_{\mu }n_{\nu }\right] 
\end{equation}
which is the corresponding three-term generalized photon propagator in the light-front gauge for the BP model.
Our result here with the three-term propagator is precisely consistent with the result that we obtained 
using the interpolation between the instant form dynamics and the light-front dynamics for the electromagnetic gauge field
\cite{JLS} and for the QED \cite{JLMS}. As discussed in Refs.\cite{JLS} and \cite{JLMS}, the last term in Eq. (\ref{3tp}) is canceled by the instantaneous interaction in the light-front dynamics so that the two-term gauge propagator given by Eq.(\ref{PPLF}) provides effectively 
the same result for the physical amplitude without involving the instantaneous interaction. As usual, the propagator of the BP model exhibits an additional simple pole at $1/a^2$.
Note further that the propagator given by Eq.(\ref{3tp})
satisfies
\begin{equation}
k^\mu P_{\mu\nu} = n^\mu P_{\mu\nu} = 0
\,,
\end{equation}
known as the double transverse property \cite{Srivastava:2000cf} in the
Maxwell case.

Curiously enough, the gauge-fixing term given by Eq.(\ref{LLF}) can be justified by imposing the two gauge conditions given by Eqs.(\ref{cond}) and (\ref{cond2}) simultaneously.
In fact, for arbitrary $r$, those two conditions together imply
\begin{equation}
0=(1+a^2\square)(\partial_\mu+r n_\mu) A^\mu
\,,
\end{equation}
and
\begin{equation}
0=(1+a^2\square)(\partial_\mu-r n_\mu) A^\mu
\,,
\end{equation}
from which we can write the generating functional as
\begin{eqnarray}
Z[j^\mu]&=&N\int DA_\mu DC^+ D{\bar C^+} DB^+
DC^- D{\bar C^-} DB^-
\exp
\left\{
                                           \vphantom{\frac{a^2\alpha}{2}\partial_\mu B^-\partial^\mu B^-}
\,\,\,
i\,S_0
\right.\nonumber\\&&\left.
+i\int d^4x 
\left[
                                                                   \vphantom{\frac{a^2\alpha}{2}\partial_\mu B^-\partial^\mu B^-}
{\bar C}^+(1+a^2\square)(\square +rn_\mu \partial^\mu) C^+
+
B^+ (1+a^2\square)(\partial^\mu  +rn^\mu )A_\mu
\right.\right.\nonumber\\&&\left.\left.
{\bar C}^-(1+a^2\square)(\square -rn_\mu \partial^\mu) C^-
+
B^- (1+a^2\square)(\partial^\mu  -rn^\mu )A_\mu
\right.\right.\nonumber\\&&\left.\left.
-{a^2\beta}\partial_\mu B^+\partial^\mu B^+
+{\beta {(B^+)}^2}
+{a^2\beta}\partial_\mu B^-\partial^\mu B^-
-{\beta {(B^-)}^2}
                                                                 \vphantom{\frac{a^2\alpha}{2}\partial_\mu B^-\partial^\mu B^-}
-j^\mu A_\mu
\right]
\right\}
\,,
\end{eqnarray}
where $B^+$ and $B^-$ denote the two Nakanishi-Lautrup fields responsible for implementing the conditions given by Eqs.(\ref{cond}) and (\ref{cond2}) while $(C^+,{\bar C}^+)$
and $(C^-,{\bar C}^-)$ are the corresponding pairs of ghost-antighost fields which come from the exponentiation of the Faddeev-Popov determinant.
Finally functionally integrating over $B^+$ and $B^-$ we get the effective action
\begin{eqnarray}
S_{eff} = S_0 +
\int d^4x
\Big\{
{\cal L}_{AL}
+{\bar C}^+(1+a^2\square)(\square +rn_\mu \partial^\mu) C^+
\\
+{\bar C}^-(1+a^2\square)(\square -rn_\mu \partial^\mu) C^-
\Big\}
\end{eqnarray}
justifying the gauge-fixing term ${\cal L}_{AL}$ in Eq.(\ref{LLF}) (for the case $r=1$).

\section{Application - The Electron Self-Energy}\label{ese}
\begin{figure}
  \centering
  \includegraphics
  [width=8cm]{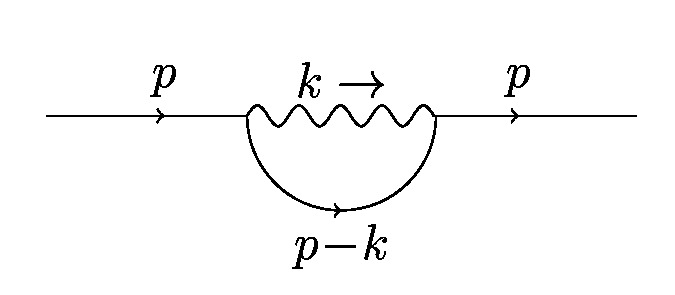}
  \caption{One-loop self-energy}
  \label{auto_energy}
\end{figure}

In this section, we discuss the second-order correction to the electron self-energy considering the propagator of the BP model for the photon.
We specifically calculate the invariant amplitude for the Feynman diagram represented in Fig.\ref{auto_energy}
which has one loop integration in the internal momentum $k$.
This effectively illustrates how BP formulation parallels with the PV regularization.

The invariant amplitude for the diagram in Fig.\ref{auto_energy} is given by
\begin{equation}\label{IA}
 {\cal M} = \bar{u}(p) \hat\Sigma(p) u(p) \, ,
\end{equation}
where $u(p)$ denotes a plane wave solution to Dirac's equation and
\begin{equation}\label{Sp}
 \hat\Sigma(p)= \int \frac{d^4 k}{(2\pi)^4}
\frac
{\gamma^\mu(\sh{p}-\sh{k}-m)\gamma^\nu}
{[(p-k)^2-m^2+i\epsilon]}
P_{\mu\nu}(k)
\end{equation}
with the photon propagator of the BP model in the Lorenz gauge given by Eq.(\ref{PPP}).
In the Feynman-t'Hooft gauge choice, $\xi=1$, and we may write more directly
\begin{equation}\label{Sp2}
 \hat\Sigma (p) =
- \int \frac{d^4 k}{(2\pi)^4}
\frac{\gamma^\mu(\sh{p}-\sh{k}-m)\gamma_\mu}
{D_{p-k}D^P_k}
\end{equation}
with the notational definitions
\begin{equation}
 D_{p-k}\equiv [ (p-k)^2 - m^2 +i\epsilon ]
\,,
\end{equation}
and
\begin{equation}
 D^P_k \equiv k^2 (1-a^2k^2)
\,.
\end{equation}

Actually, even for an arbitrary value of $\xi$, the calculation of the simpler expression (\ref{Sp2}) is sufficient
to obtain the full invariant amplitude in Eq.(\ref{IA}) for the original $\hat\Sigma (p)$ in Eq.(\ref{Sp}).
This comes from the fact that the gauge term originated from Eq.(\ref{PPP}) and inserted into Eq.(\ref{Sp}),
\begin{equation}\nonumber
 (\xi-1)\frac{k_\mu k_\nu}{k^2}
\end{equation}
does not contribute to (\ref{IA}) as can be seen from
\begin{eqnarray}
 \gamma^\mu
 \frac{1}{\sh{p}-\sh{k}-m}
 \gamma^\nu k_\mu k_\nu u(p)
 &=&
 \left[
 (\sh{p}-m)-(\sh{p}-\sh{k}-m)
 \right]
 \frac{1}{\sh{p}-\sh{k}-m}
 \sh{k} u(p)
 \nonumber\\
 &=&-\sh{k} u(p).
\end{eqnarray}
Since the remaining denominator is even in $k$ this term vanishes after the momentum integration in Eq.(\ref{Sp}).
By using standard gamma matrix properties, Eq.(\ref{Sp2}) can be further simplified to
\begin{equation}
 \hat \Sigma(p) = 2 \int \frac{d^4 k}{(2\pi)^4}
\frac{[(\sh{p}-\sh{k})-2m]}{D_{p-k}D^P_k}
\,.
\end{equation}

Next, we split the denominator for the photon propagator as the sum
\begin{equation}
 \frac{1}{D^P_k}=
\frac{1}{k^2}-\frac{1}{k^2-\Lambda^2}
\end{equation}
where $\Lambda\equiv 1/a$
and define further
\begin{equation}
 D^\Lambda_k \equiv k^2-\Lambda^2+i\epsilon
\end{equation}
to obtain the handy relation
\begin{equation}
 \hat\Sigma(p)=2 \int \frac{d^4 k}{(2\pi)^4}
\left[
\frac{(\sh{p}-\sh{k})-2m}
{D_{p-k}D^0_k}
-
\frac{(\sh{p}-\sh{k})-2m}
{D_{p-k}D^\Lambda_k}
\right]
\,.
\end{equation}
Applying this operator into the Dirac equation free plane wave solution $u(p)$, as necessary for
plugging into Eq.(\ref{IA}), we may use Dirac's equation to get
\begin{equation}\label{Sspl}
 \hat\Sigma(p)u(p)=
2
\left(
\hat\Sigma_\Lambda(p)
-\hat\Sigma_0(p)
\right)
u(p)
\end{equation}
with
\begin{equation}
 \hat\Sigma_\Lambda(p)
\equiv
\int \frac{d^4 k}{(2\pi)^4}
\frac{\sh{k}+m}
{D_{p-k}D^\Lambda_k}
\,.
\end{equation}
Using the well known Feynman parametrization technique, we may
express the inverse of the product $D_{p-k}D_k^\Lambda$ as an integral over a dummy real variable
$x$ and
 write
\begin{equation}
 \hat\Sigma_\Lambda(p)
=
\int_0^1 dx
\int \frac{d^4 k}{(2\pi)^4}
\frac{\sh{k}+m}
{
\left[ k'^2 - M(x,p,\Lambda)^2 + i\epsilon \right]^2
}
\end{equation}
with
\begin{equation}
 k'\equiv k - xp
\end{equation}
and
\begin{equation}
 M(x,p,\Lambda)^2 \equiv (1-x)\Lambda^2 + x^2p^2
\,.
\end{equation}
After changing the momentum integration variable to $k'$, renaming it back to $k$, and using again Dirac's equation
we may write
\begin{equation}
  \hat\Sigma_\Lambda(p)
\,u(p)
=
\int_0^1 dx
\int \frac{d^4 k}{(2\pi)^4}
\frac{m(1+x)}
{
\left[ k^2 - M(x,p,\Lambda)^2 + i\epsilon \right]^2
}
\,u(p)
\,.
\end{equation}
Subsequently, we consider dimensional regularization and define the quantity
\begin{equation}\label{dr}
  \hat\Sigma_\Lambda^{(\epsilon)}(p)
=
m\mu^{4-D}
\int_0^1 dx
\int \frac{d^D k}{(2\pi)^D}
\frac{1+x}
{
\left[ k'^2 - M(x,p,\Lambda)^2 + i\epsilon \right]^2
}
\end{equation}
where $D=4-2\epsilon$ denotes a general real dimension, with $\epsilon$ infinitesimal, which
reproduces Eq.(\ref{dr}) when multiplied by $u(p)$ in the limit $D\rightarrow 4$.

Now, from
\begin{eqnarray}
 \mu^{4-D}\int \frac{d^D k}{(2\pi)^D} \frac{1}{(k^2-M^2)^2}
 &=&
 \frac{i}{16\pi^2}\left(\frac{4\pi\mu}{M^2}\right)^\epsilon \Gamma(\epsilon)\nonumber\\
 &=&
 \frac{i}{16\pi^2}
 \left[
 \frac{1}{\epsilon}-\gamma+\log \frac{4\pi\mu^2}{M^2} + o(\epsilon) 
 \right]\, ,
\end{eqnarray}
we get 
\begin{equation}
\hat\Sigma_\Lambda^{(\epsilon)}(p)
=
 \frac{i}{16\pi^2}
\int_0^1 dx(1-x)
\left[
\frac{1}{\epsilon}-\gamma+\log \frac{4\pi\mu^2}{(1-x)\Lambda^2+x^2m^2}
\right]
+o(\epsilon)
\,.
\end{equation}

This result diverges for $\epsilon\rightarrow 0$, but now we may come back to Eq.(\ref{Sspl}), where the divergent parts for
$\hat\Sigma_\Lambda$ and $\hat\Sigma_0$ cancel, to get the finite result given by
\begin{equation}
\label{final-result}
 \hat\Sigma (p) =
 \frac{im}{8\pi^2}
\int dx (1+x)
\log\frac{x^2m^2}{x^2m^2+(1-x)\Lambda^2} \, .
\end{equation}
Naturally, the invariant amplitude given by Eq.(\ref{IA}) must be gauge independent.  Instead of Eq.(\ref{PPP}), even if one uses Eq. (\ref{PPLF}) for the light-front gauge photon propagator, the same result must be achieved.  Indeed, it is easy to show that the second term in Eq.(\ref{PPLF}) does not contribute.
Consider the identity
\begin{equation}\label{ILF}
 \gamma^\mu \frac{1}{\sh{p}-\sh{k}-m}\gamma^\nu
\left[
\frac{k_\mu n_\nu + k_\nu n_\mu}{k\cdot n}
\right]
=
\left[
\sh{k}\frac{1}{\sh{p}-\sh{k}-m}\sh{n}
+
\sh{n}\frac{1}{\sh{p}-\sh{k}-m}\sh{k}
\right]
\frac{1}{n\cdot k}
\end{equation}
Multiplying by $\bar{u}(p)$ from the left and $u(p)$ on the right, as demanded by Eq.(\ref{IA}), and using once more Dirac's equation
the identity given by Eq.(\ref{ILF}) leads to
\begin{eqnarray}
\bar{u}(p)
 \left\{
\left[(\sh{p}-m)-(\sh{p}-\sh{k}-m)\right]\frac{1}{\sh{p}-\sh{k}-m}\sh{n}
\right.&&\nonumber\\\left.
+\sh{n}\frac{1}{\sh{p}-\sh{k}-m}\left[(\sh{p}-m)-(\sh{p}-\sh{k}-m)\right]
 \right\}
\frac{1}{n\cdot k}
u(p)
&
=&
 -2\bar{u}(p)\frac{\sh{n}}{n\cdot k} u(p)
\end{eqnarray}
which, being odd in $k$, amounts to zero after momentum integration.
Even if we use the three-term propagator given by Eq.(\ref{3tp}), 
the last term in Eq. (\ref{3tp}) is canceled by the instantaneous interaction in the light-front dynamics~\cite{JLS,JLMS}
and thus the result is identical to Eq.(\ref{final-result}).
It shows the gauge independence of the invariant amplitude given by Eq.(\ref{IA}) and illustrates how BP formulation parallels with the PV regularization.

\section{Conclusion and Discussion}\label{con}
In this work, we demonstrated that the BP model solution for a point charge in electrodynamics corresponds to an ordinary electrodynamic solution for a specific charge distribution given by Eq.(\ref{rho2}). 
Motivated by this possible reinterpretation of BP model solution in electrodynamics, we further elucidate the BP model 
as a natural way of providing Pauli-Villars regularization in ordinary QED.  

We have pursued the gauge fixing of BP's generalized electromagnetism in three distinct ways.
Specifically we analyzed the standard Lorenz, axial and light-front gauge fixings.
For all considered cases, we have achieved a clean and neat generalized photon propagator depending
on two parameters, namely the free gauge parameter and BP's length dimensional parameter.
As a general result, we have shown the propagator of the BP model can have the same structure
of the usual Maxwell case with only an additional pole at $k^2=1/a^2$.
Note that the common multiplicative factor in the propagator can be split as 
\begin{equation}
\frac{1}{k^2(1-a^2k^2)} = \frac{1}{k^2}-\frac{1}{k^2-(1/a)^2}.
\end{equation}
Thus, it can be interpreted as the sum of two propagators describing a massless photon and a massive one corresponding
to the Pauli-Villars regularization.
We have shown that the different gauge fixings considered do not invalidate this appealing interpretation.
\\

\textbf{Acknowledgments:}\\
This work was supported by the U.S. Department of Energy (Grant No. DE-FG02-03ER41260).
A.T.S. wishes to thank the kind hospitality of Physics Department, North Carolina State University, Raleigh, NC and
acknowledges research grant in the earlier part of this work from Fapesp 2014/20892-2.
J.H.O.S. thanks for the hospitality of North Carolina State University,
Raleigh, NC which provided facilities for the completion of this work and
thanks the financial support of AUXPE-FAPESB-3336/2014/Processo no: 23038.007210/2014-19 and CNPq.

\appendix

\section{The $\bar\psi(a,b,r)$ Integral}
\label{A}

In this appendix, we compute the integral
\begin{equation*}
\bar\psi(a,b,r)\equiv   
\frac{1}{4\pi b^2}\int \,d\tau'\,\,
\frac{ \exp\left[-\frac{r'}{b}-\frac{|\mathbf{r}-\mathbf{r}'|}{a}\right]}{r'|\mathbf{r}-\mathbf{r}'|}\,\,.
\tag{\ref{barpsiabr}}
\end{equation*}
which was used in Section {\bf \ref{pge}} to obtain the double BP potential.
By using spherical coordinates
\begin{equation}
d\tau'= r'^2\sin\theta \,d\phi \,d\theta \,dr'
\end{equation}
we integrate in the azimuthal variable $\phi$ and rewrite (\ref{barpsiabr}) as
\begin{equation}\label{intbarpsi}
2b^2\bar\psi(a,b,r) =\int_0^\infty\,dr'\,r'e^{-r'/b}\,I(r,r')\,,
\end{equation}
with
\begin{equation}
I(r,r')\equiv\int_0^\pi\,\frac{d\theta\,\sin\theta\,\, e^{-|\mathbf{r}-\mathbf{r}'|/a}}{|\mathbf{r}-\mathbf{r}'|}\,.
\end{equation}
The integration in $\theta$ from $0$ to $\pi$ gives
\begin{equation}
I(r,r') = \frac{a}{r\,r'}
\left[
e^{-\frac{|{r}-{r}'|}{a}}-e^{-\frac{|{r}+{r}'|}{a}}
\right]
\end{equation}
which substituted into (\ref{intbarpsi}) leads to
\begin{eqnarray}
\frac{2b^2 r}{a}\bar\psi(a,b,r) &=&\int_0^\infty\,dr'
e^{-r'\over b}\left[
e^{-\frac{|{r}-{r}'|}{a}}-e^{-\frac{|{r}+{r}'|}{a}}\right]
\nonumber\\
&=&-e^{-{r/a}}
\int_0^\infty\, dr'\,e^{-r'\left(\frac{a+b}{ab}\right)}
+e^{-{r/a}}
\int_0^r \, dr'\,e^{-r'\left(\frac{a-b}{ab}\right)}
+e^{{r/a}}
\int_r^\infty\, dr'\,e^{-r'\left(\frac{a+b}{ab}\right)}
\nonumber\\&=&
e^{-\frac{r}{a}}\left(\frac{2ab^2}{a^2-b^2}\right)
-
e^{-\frac{r}{b}}\left(\frac{2ab^2}{a^2-b^2}\right)
\,.
\end{eqnarray}
Cancelling the $2b^2$ common factor and regrouping terms we obtain
\begin{equation*}
\bar\psi(a,b,r)=
-\frac{e^{-{r\over b}} - e^{-\frac{r}{a}}}{r[1-(b/a)^2]}
\tag{\ref{barpsiabr2}}
\end{equation*}
which represents the result used in the main text to be substituted into equation (\ref{psiabr2}).

\end{document}